%% file: Interspeech_2024_remi_uro.tex
\documentclass{Interspeech2024}

\usepackage{multirow}
\usepackage{svg}

\interspeechcameraready

\title{Detecting the terminality of speech-turn boundary for spoken interactions in French TV and Radio content}

\name[affiliation={1,2}]{Rémi}{Uro}
\name[affiliation={3}]{Marie}{Tahon}
\name[affiliation={1}]{David}{Doukhan}
\name[affiliation={3}]{Antoine}{Laurent}
\name[affiliation={2}]{Albert}{Rilliard}

\address{
  $^1$French National Institute of Audiovisual (INA), Paris, France.
  $^2$Université Paris Saclay, CNRS, LISN, Orsay, France.
  $^3$LIUM, Le Mans Université, France}
\email{ruro@ina.fr, marie.tahon@univ-lemans.fr, ddoukhan@ina.fr, antoine.laurent@univ-lemans.fr, albert.rilliard@lisn.fr}


\newcommand{\red}[1]{\textcolor{red}{#1}}

\begin{document}

\maketitle

\begin{abstract}

Transition Relevance Places are defined as the end of an utterance where the interlocutor may take the floor without interrupting the current speaker --i.e., a place where the turn is terminal. Analyzing turn terminality is useful to study the dynamic of turn-taking in spontaneous conversations.
This paper presents an automatic classification of spoken utterances as Terminal or Non-Terminal in multi-speaker settings. We compared audio, text, and fusions of both approaches on a French corpus of TV and Radio extracts annotated with turn-terminality information at each speaker change. Our models are based on pre-trained self-supervised representations. We report results for different fusion strategies and varying context sizes. This study also questions the problem of performance variability by analyzing the differences in results for multiple training runs with random initialization.
The measured accuracy would allow the use of these models for large-scale analysis of turn-taking.

\end{abstract}

\section{Introduction}

Turn-taking dynamics \cite{Couper-Kuhlen_1993} and its relations to interruptions are important factors in describing broadcast interactions, as they hint at perceived power and dominance in a conversation \cite{Zimmerman_West_1996, Beattie_1982}.
\cite{Sacks_Schegloff_Jefferson_1974} introduced the notion of Transition Relevance Places (TRPs), which occur at the end of turn-construction units and allow for smooth transitions between speakers.
TRPs are important because speakers can anticipate their occurrences in order to plan a potential turn-taking, optimizing the floor transfer offset \cite{Levinson_2016}. 
Automatically detecting TRPs would allow for large-scale analysis of complex interruptions-related phenomena. 
As for other human science objects, allowing large datasets describing turn-tacking dynamics to be produced and analyzed is an important outcome for society.
While \cite{Lebourdais_Tahon_Laurent_Meignier_Larcher_2022} focuses on overlapping speech segments to study interruptions following a classification schema defined by \cite{addadecker2008} for French political interviews, interruptions may also occur without overlapping speech \cite{levinson_pragmatics_1983}.
A better understanding of turn-taking also has implications for the development of human-machine interactions \cite{Hara_Inoue_Takanashi_Kawahara_2019, Skantze_2021}.
\cite{Ondas_Pleva_Bacikova_2023} has annotated TRPs in 8~hours of Slovak TV discussions and reports a binary classification accuracy of 94.4\% with an ensemble model using fundamental frequency (F0) and intensity curves on chunks of 1.5~s.
\cite{Skantze_2017,Roddy_Skantze_Harte_2018} propose an LSTM-based architecture using acoustic and linguistic features on English spontaneous dialogues to predict whether a speaker change would occur in the three following seconds. \cite{Ekstedt_Skantze_2020} presents a Transformer-based approach to the same task based on textual information only.
\cite{Hara_Inoue_Takanashi_Kawahara_2019} proposed a method for turn-taking prediction in spoken dialog systems. They report substantial inter-rater agreements for annotating TRPs in different acted dialogue scenarios. The proposed LSTM architecture using acoustic and linguistic features achieves binary accuracies between 79.3 and 89.5.

The work presented here focuses on media broadcasts, especially for shows proposing multispeaker interactions. This choice was made to investigate spontaneous spoken dialogs and to allow collaboration with sociologists analyzing roles and behavior in media. 
We present an automatic classification of spoken utterances as Terminal (i.e., ending with a TRP) or Non-Terminal (not ending with a TRP). The corpus was based on French TV and radio content, specifically multi-party conversation representing spontaneous interactions with various \textit{levels of control} \cite{Wagner_Trouvain_Zimmerer_2015}. 
An existing corpus annotated with TRPs~\cite{urolrec24} was used to train multimodal models based on the Wav2Vec2~\cite{Baevski_Zhou_Mohamed_Auli_2020} and FlauBERT~\cite{Le_Vial_Frej_Segonne_Coavoux_Lecouteux_Allauzen_Crabbé_Besacier_Schwab_2020} pre-trained models. We report results comparing different fusion strategies and context sizes, evaluate the relative role of the different information modalities (lexical vs. acoustic) with respect to the length of the segments used for the inference, and discuss the relevance of these findings for spoken interaction modeling.

\section{Method}

\subsection{Data}

This study is based on an annotated corpus of multi-party interactions composed of French TV and Radio broadcasts from 1998 to 2015~\cite{urolrec24} from ALLIES corpus~\cite{tahon2024allies}.
Audio chunks corresponding to zones of turn changes were extracted from the complete show to encompass (1) the last speech \textit{segment}\footnote{defined by the Allies terminology as “sequences containing complete words which are syntactically and semantically coherent”~\cite{larcher:hal-03262914}} before the turn change and (2) the first segment after the change. The segment (1) comes before the turn change and does not include potential overlapped speech, while segment (2) may contain an overlap. 
These audio chunks were annotated regarding the terminality of the initial speaker's turn (terminal or not) and the second speaker's turn-taking category (interruption, backchannel, or smooth change).
One annotator paid for this work (a student trained in linguistics) made the annotations. Two additional persons also annotated a subset of 338 samples to obtain an inter-rater agreement. A substantial agreement of 0.75 using Fleiss' Kappa~\cite{Fleiss_1971} was observed for the Terminality annotation used in this paper. 
The work presented here focuses solely on the parts (1), preceding the speaker change, to avoid relying on information linked to the second speaker's intervention.
This was decided because we aim to produce a model that classifies speech turn in relation to turn-taking management and the anticipation of a given speaker floor taking \cite{Levinson_2016}, thus without knowing what happened later during the dialogue. 
We selected the segment (1) occurring before the speaker changes, which constitute the "samples" that are used in the remaining of this study.
These samples amounted to a total of 1954 speaker changes annotated with TRP information with 128 different speakers. Table~\ref{tab:corpus} shows the number of samples and their duration characteristics for each Terminality class.



\begin{table}[th]
    \centering
    \begin{tabular}{|l|r|r|r|r|r|}
         \hline
         & Nb & $\leq$0.5 & 0.5$<$x$\leq$1 & 1$<$x$\leq$2 & $>$2\\
         \hline
         Term. & 839 & 13\% & 27\% & 35\% & 25\%\\
         Non Term. & 1115 & 11\% & 19\% & 23\% & 47\%\\
         \hline
    \end{tabular}
    \caption{Number of samples and percentage of samples in different duration intervals (in seconds) for the Terminal and Non-Terminal classes}
    \label{tab:corpus}
\end{table}

Table~\ref{tab:shows} lists the names of the different shows used to select the samples
and the number and total duration of the annotated samples from each show.


\begin{table}[t]
    \centering
    \begin{tabular}{lrr}
\toprule
Show & Samples & Dur. (s)\\
\midrule
BFMStory (BS) & 200 & 853.34\\
CaVousRegarde (CR) & 269 & 969.84\\
CultureEtVous (CV) & 7 & 8.69\\
DEBATE (D) & 128 & 266.22\\
EntreLesLignes(EL) & 307 & 1179.53\\
LaPlaceDuVillage (PV) & 770 & 1316.76\\
PileEtFace (PF) & 150 & 573.33\\
PlaneteShowbiz (PS) & 10 & 13.94\\
TopQuestions (TQ) & 5 & 21.88\\
fm (FM) & 108 & 254.25\\
\bottomrule
\end{tabular}
    \caption{Number samples from each show, and their total duration}
    \label{tab:shows}
\end{table}

Two settings for the chunks used as input of the model were tested: (1) the variable-size samples defined above, that comprises a complete "segment" as defined in the Allies corpus  (thereafter \textit{ref}); or (2) a fixed size chunk that correspond to the three seconds before the turn taking event occurring at the end of segment 1, and that could have been obtained using an automatic diarization  (thereafter \textit{3s}; these may include other speakers turns at the beginning, or incomplete turn). 
This was done to evaluate the possibility to apply a fully automatic approach.

For the fixed-sized chunks, other sizes have been tested in preliminary evaluations (2 and 5 seconds) with similar results. Three-second chunks were kept specifically to investigate if the model would work even if provided with parts of previous turns.

\subsection{Data preprocessing}

For each sample, we extracted the relevant audio waveform, either the annotated segment for \textit{ref} or the last 3 seconds of the recordings before the speaker change for \textit{3s}. 
All samples were then automatically transcribed using Whisper~\cite{Radford_Kim_Xu_Brockman_McLeavey_Sutskever_2022}. 
To evaluate the model in fully automatic processing pipeline conditions, we kept the Whisper transcriptions as is, even if easily detectable errors occurred (e.g., "Subtitles generated by\ldots"). 
Manual transcriptions for the \textit{ref} setting were also available from the Allies reference transcriptions and are used for comparison.

\subsection{Model architectures}

\begin{figure}[th]
    \centering
    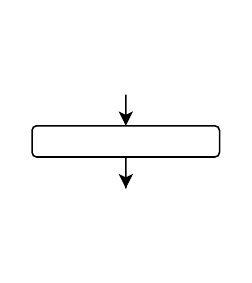
    \caption{Single modality model architecture}
    \label{fig:model_single}
\end{figure}

\begin{figure}[th]
    \centering
    \def\svgwidth{.5\textwidth}
    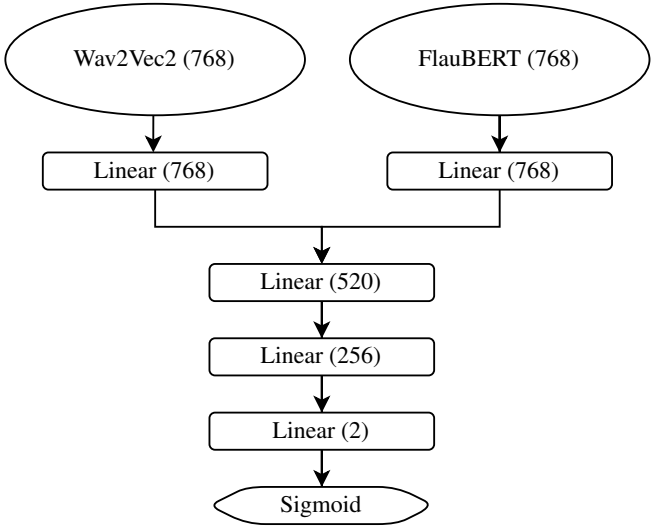
    \caption{Early fusion model architecture}
    \label{fig:model_both}
\end{figure}

A total of 5 prediction models are proposed, operating on raw audio and/or textual transcriptions.
The models are based on the pre-trained self-supervised models \texttt{wav2vec2-base}~\cite{Baevski_Zhou_Mohamed_Auli_2020} for the audio and \texttt{flaubert\_base}~\cite{Le_Vial_Frej_Segonne_Coavoux_Lecouteux_Allauzen_Crabbé_Besacier_Schwab_2020} for the text. They both output a classification token (CLS) of size 768 from variable-size audio or text inputs.
Text-only (\textbf{TO}) and Audio-only (\textbf{AO}) consist of three linear layers after extracting the CLS token of the pre-trained model (see Figure~\ref{fig:model_single}). 
Three additional fusion processes inspired by \cite{Tahon_Macary_Estève_Luzzati_2021} are defined: 
Early Fusion (\textbf{EF}), with one linear layer after each pre-trained model before concatenation and another three linear layers (see Figure~\ref{fig:model_both}); Late Fusion (\textbf{LF}) with one linear layer on the concatenation of the TO and AO outputs; and Average Fusion (\textbf{AF}) which takes the average of the logits of TO and AO. A dropout of 0.30 was applied between each linear layer during training.

\section{Results}

For better convenience, the model's input used during training or at inference time will be referred to using the following symbols: \textbf{ref\_auto}: reference speaker segmentation and their automatic speech transcription;
\textbf{ref\_man}: reference speaker segmentation and their manual speech transcription;
\textbf{3s\_auto} fixed 3-second duration excerpts and their automatic speech transcription.


\subsection{Evaluation protocol}

Models were trained and cross-tested for each train and inference input combination (\textit{ref\_auto}, \textit{ref\_man} and \textit{3s\_auto}), resulting in nine evaluation configurations for each architecture.
Each model was evaluated using K-Fold grouped by show from which the examples were extracted. 
This was done to prevent having samples of the same show in the training and testing sets and to allow for a comparison of the performance degradation associated with materials of different natures (e.g., different types of shows).
A random split of 30\% of the training set was used as a validation set. 
A patience of five epochs was used on validation accuracy.
This process was run 10 times to estimate the variability due to random initialization, allowing for more reliable accuracy measures (mean and confidence intervals over the 10 runs are given).
The training was done on an RTX 4090 GPU, and it took about 80 hours to train the 1500 models for the whole experiment.
In order to investigate the significance of accuracy variation and the impact of the different settings, a linear mixed model was fit on the accuracy of these model architectures across training and test sets, with the samples nested in the show as random factors, as in equation \ref{eq:lmm}.
\begin{equation}
accuracy \sim model \ast (test + train) + (1 | show/sample) 
\label{eq:lmm}
\end{equation}
where the $accuracy$ is explained by the $model$ architecture and its interactions with the $test$ and $train$ settings as fixed factors, and the $sample$ nested in $show$ as random factors. 
The model was fitted using R's \textit{lme4} library \cite{rcoreteam, Bates_2015}.
The mean and confidence interval fitted by the models are plotted in Figure~\ref{fig:effects}.
A post hoc comparison of the models' accuracies (using Bonferroni correction) for each level of test and train settings was done.

\subsection{Variation across models}

\begin{figure}[!h]
    \centering
    \includegraphics[width=.5\textwidth]{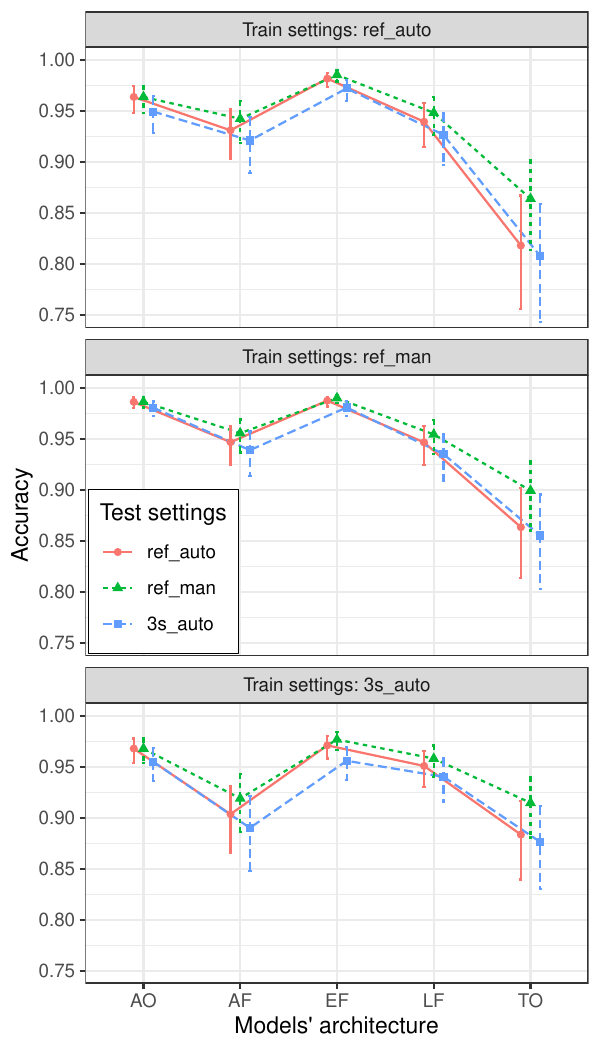}
    \caption{Mean accuracies for each model's Test settings (lines) with their confidence intervals, for each Train setting (subplots), for the different model architectures (x-axis).}
    \label{fig:effects}
\end{figure}

All models achieved an average accuracy score above 0.85 as can be seen on Figure~\ref{fig:effects}.
This indicates that the proposed approach reached coherent and qualitative results.
Figure~\ref{fig:effects} shows the mean accuracies with their confidence intervals for each combination of model architecture, training set, and testing set, as estimated by the linear mixed model across all random initialization runs. 
We observed that the Text Only approach performs significantly worse than the other models. 
Then comes the Late and Average Fusion models, which significantly outperform the TO approach and achieved comparable performances (in terms of accuracy) but are significantly outperformed by the Early Fusion and Audio Only approaches. 
AO and EF achieve comparable performances, hinting that the audio signal carries an important share of information for this task, including non-linguistic cues. 
These trends are similar for each training and testing setting.
The training settings do not seem to have much impact, all models achieved similar accuracies across training configurations, with the ones trained on the \textit{ref\_man} performing slightly better.

Tables~\ref{tab:acc_by_show_man} and~\ref{tab:acc_by_show_3s} report the mean accuracy for each show, respectively, with the \textit{ref\_man} and \textit{3s} models for the different test settings. 
The EF model performs slightly better than the AO for most shows. 
Meanwhile, considering the confidence intervals obtained from the ten random runs, the difference does not reach significance.
The worst accuracies for both training settings are observed on the \textit{PlanetShowbiz} show, which contains fewer shows (10) and displays an interaction style that differs from the others, with presenters often reading prepared scripts. 
Using the reference segmentation leads to better results than using a fixed size of three seconds, but a fully automatic approach still reaches accuracies over 90\%.

\begin{table}[th]
    \centering
        \fontsize{9}{9}\selectfont
\begin{tabular}{l|rr|rrr}
\toprule
Show & \multicolumn{2}{c|}{AO} & \multicolumn{3}{c}{EF} \\

 & ref & 3s & ref & ref\_man & 3s \\
\midrule

(BS) & 97.86 & 96.21 & 98.14 & \textbf{98.21} & 95.79 \\
(CR) & \textbf{96.49} & 94.74 & 95.27 & 96.02 & 93.73 \\
(CV) & \textbf{97.96} & 87.76 & 95.92 & 95.92 & 89.80 \\
(D)  & \textbf{96.54} & 94.42 & 96.09 & \textbf{96.54} & 94.20 \\
(EL) & \textbf{97.16} & 94.23 & 93.62 & \textbf{97.16} & 88.83 \\
(PV) & 91.41 & 87.25 & 93.75 & \textbf{95.84} & 90.95 \\
(PF) & 98.76 & 97.43 & 95.90 & \textbf{99.33} & 93.81 \\
(PS) & \textbf{84.29} & 77.14 & 82.86 & \textbf{84.29} & 74.29 \\
(TQ) & 91.43 & 91.43 & \textbf{100.00} & \textbf{100.00} & \textbf{100.00} \\
(FM) & \textbf{96.16} & 94.18 & \textbf{96.16} & 95.77 & 95.24 \\
\midrule
\textbf{Mean} & 94.80 & 91.48 & 94.77 & \textbf{95.90} & 91.66 \\
\bottomrule
\end{tabular}
    \caption{Mean accuracies for each show with the AO and EF model trained in the \textit{ref\_man} setting}
    \label{tab:acc_by_show_man}
\end{table}

\begin{table}[]
    \centering
    \fontsize{9}{9}\selectfont
    \begin{tabular}{l|rr|rrr}
\toprule
Show & \multicolumn{2}{c|}{AO} & \multicolumn{3}{c}{EF} \\

 & ref & 3s & ref & ref\_man & 3s \\
\midrule
(BS) & 97.86 & 97.64 & \textbf{98.29} & \textbf{98.29} & 97.50 \\
(CR) & 96.49 & 96.23 & \textbf{96.81} & \textbf{96.81} & 95.75 \\
(CV) & \textbf{93.88} & \textbf{93.88} & \textbf{93.88} & \textbf{93.88} & 91.84 \\
(D) & 96.76 & \textbf{97.66} & 95.87 & 95.76 & 96.76 \\
(EL) & 97.86 & 95.86 & \textbf{97.95} & 97.91 & 96.04 \\
(PV) & 88.09 & 85.77 & 96.07 & \textbf{96.38} & 93.65 \\
(PF) & 98.86 & 98.86 & 98.48 & \textbf{98.95} & 97.81 \\
(PS) & \textbf{87.14} & 81.43 & \textbf{87.14} & \textbf{87.14} & 81.43 \\
(TQ) & \textbf{94.29} & 88.57 & \textbf{94.29} & \textbf{94.29} & 91.43 \\
(FM) & 96.16 & 94.58 & \textbf{96.30} & \textbf{96.30} & 95.24 \\
\midrule
\textbf{Mean} & 94.74 & 93.05 & 95.50 & \textbf{95.56} & 93.75\\
\bottomrule
\end{tabular}
    \caption{Mean accuracies for each show with the AO and EF model trained in the \textit{3s} setting}
    \label{tab:acc_by_show_3s}
\end{table}

\begin{figure}[!ht]
    \centering
    \includegraphics[width=.4\textwidth]{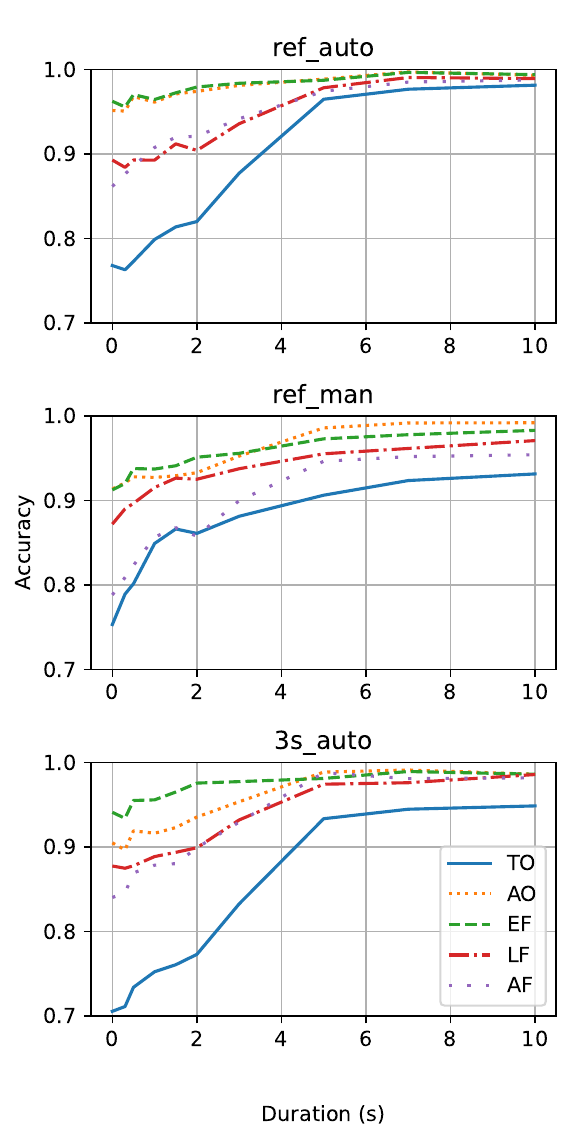}
    \caption{Mean accuracies as a function of the duration of the testing sample for the different three training settings.
    }
    \label{fig:accs_by_dur}
\end{figure}

Figure~\ref{fig:accs_by_dur} represents the mean accuracy for each model architecture depending on the training setting and the duration of the speech segment in the training sample. 
We observe that textual information on shorter segments is less reliable than the audio signal, but tends to allow similar performances on longer segments. 
This is especially true with the automatic settings, as very short segments may not include enough data for a reliable transcription.

\section{Discussion \& conclusions}

We compared different models and data settings to classify the terminality status of speech samples extracted from spontaneous interactions in media archives.
While the proposed approaches achieved highly encouraging performances, these models are based on a corpus of about 2000 audio samples from 10 different shows with limited types of interactions and have not yet been tested on other corpora.
One important takeaway of this study is that the \textit{Wav2Vec2} representation alone seems sufficient for classifying Transition Relevance Places, as Early Fusion and Audio Only models achieve comparable performances. However, the AO model is smaller and does not require transcription as a preprocessing step, making it less resource-demanding and thus more easily usable in real-life settings.
It also potentially has fewer language dependencies than a model (partially) based on text, albeit we didn't test this.
The Text Only models also achieve satisfying performances, especially on longer speech turns, showing that lexical information is relevant for this task but, as pointed out by \cite{Ekstedt_Skantze_2020}, more work on language modeling, specifically focused on the varying levels of boundaries (e.g., \cite{barbosa2018spontaneous}), would be important.
Comparing different settings for data preprocessing (manual segmentation vs. fixed-size window; manual vs. automatic transcription) shows that the proposed models are fairly invariant to the preprocessing setting, hinting that they could be used in a fully automatic setting. Applying automatic diarization may also help to reduce the gap between manual and automatic settings at inference.
Testing the proposed models on other languages and interaction types could prove useful in better understanding the differences in turn-taking dynamics depending on context.
The performances that the proposed models achieved are coherent with the most similar works in the literature~\cite{Ondas_Pleva_Bacikova_2023} which is based on prosodic features alone.
We believe rendering our dataset available will improve evaluation homogeneity among similar work.
We think the notion of terminal speech turn is important for better-describing interactions, and the results we obtained here encourage us to assert that it is a notion that can be automatically predicted from spontaneous speech.
This work can have applications for large-scale semi-automatic media analysis. The annotation of interruptions in the media is a subjective task, with multiple studies not necessarily obtaining the same conclusions~\cite{Sandré_2009, Constantin_de_Chanay_Kerbrat-Orecchioni_2010}. The proposed system would allow for a reproducible method for automatically highlighting places of interest in a long document and thus enable researchers to focus on the speaker, specifically changes that happen outside of a detected TRP on French TV and Radio content.
This corpus is also annotated with backchannels and interruption information, future works will focus on the classification of turn-taking with regard to these phenomena.

\section{Availability}

Code and datasets used in this paper are available at \url{https://github.com/ina-foss/termClassif}.

\section{Acknowledgements}

This work has been partially funded by the French National Research Agency and the German DFG (GEM project - ANR-19-CE38-0012 and CLD 2025 - ANR-19-CE38-0015).

\bibliographystyle{IEEEtran}
\bibliography{mybib}

\end{document}

%% file: model_single_svg-tex.pdf_tex
\begingroup%
  \makeatletter%
  \providecommand\color[2][]{%
    \errmessage{(Inkscape) Color is used for the text in Inkscape, but the package 'color.sty' is not loaded}%
    \renewcommand\color[2][]{}%
  }%
  \providecommand\transparent[1]{%
    \errmessage{(Inkscape) Transparency is used (non-zero) for the text in Inkscape, but the package 'transparent.sty' is not loaded}%
    \renewcommand\transparent[1]{}%
  }%
  \providecommand\rotatebox[2]{#2}%
  \newcommand*\fsize{\dimexpr\f@size pt\relax}%
  \newcommand*\lineheight[1]{\fontsize{\fsize}{#1\fsize}\selectfont}%
  \ifx\svgwidth\undefined%
    \setlength{\unitlength}{120.75bp}%
    \ifx\svgscale\undefined%
      \relax%
    \else%
      \setlength{\unitlength}{\unitlength * \real{\svgscale}}%
    \fi%
  \else%
    \setlength{\unitlength}{\svgwidth}%
  \fi%
  \global\let\svgwidth\undefined%
  \global\let\svgscale\undefined%
  \makeatother%
  \begin{picture}(1,1.1242236)%
    \lineheight{1}%
    \setlength\tabcolsep{0pt}%
    \put(0,0){\includegraphics[width=\unitlength,page=1]{model_single_svg-tex.pdf}}%
    \put(0.49689441,0.54037267){\color[rgb]{0,0,0}\makebox(0,0)[t]{\lineheight{1.25}\smash{\begin{tabular}[t]{c}Linear (520)\end{tabular}}}}%
    \put(0,0){\includegraphics[width=\unitlength,page=2]{model_single_svg-tex.pdf}}%
    \put(0.49689441,0.29192547){\color[rgb]{0,0,0}\makebox(0,0)[t]{\lineheight{1.25}\smash{\begin{tabular}[t]{c}Linear (256)\end{tabular}}}}%
    \put(0,0){\includegraphics[width=\unitlength,page=3]{model_single_svg-tex.pdf}}%
    \put(0.49689441,0.04347826){\color[rgb]{0,0,0}\makebox(0,0)[t]{\lineheight{1.25}\smash{\begin{tabular}[t]{c}Linear (2)\end{tabular}}}}%
    \put(0,0){\includegraphics[width=\unitlength,page=4]{model_single_svg-tex.pdf}}%
    \put(0.49689441,0.91304348){\color[rgb]{0,0,0}\makebox(0,0)[t]{\lineheight{1.25}\smash{\begin{tabular}[t]{c}Pretrained (768)\end{tabular}}}}%
  \end{picture}%
\endgroup%

%% file: model_both_svg-tex.pdf_tex
\begingroup%
  \makeatletter%
  \providecommand\color[2][]{%
    \errmessage{(Inkscape) Color is used for the text in Inkscape, but the package 'color.sty' is not loaded}%
    \renewcommand\color[2][]{}%
  }%
  \providecommand\transparent[1]{%
    \errmessage{(Inkscape) Transparency is used (non-zero) for the text in Inkscape, but the package 'transparent.sty' is not loaded}%
    \renewcommand\transparent[1]{}%
  }%
  \providecommand\rotatebox[2]{#2}%
  \newcommand*\fsize{\dimexpr\f@size pt\relax}%
  \newcommand*\lineheight[1]{\fontsize{\fsize}{#1\fsize}\selectfont}%
  \ifx\svgwidth\undefined%
    \setlength{\unitlength}{258.75bp}%
    \ifx\svgscale\undefined%
      \relax%
    \else%
      \setlength{\unitlength}{\unitlength * \real{\svgscale}}%
    \fi%
  \else%
    \setlength{\unitlength}{\svgwidth}%
  \fi%
  \global\let\svgwidth\undefined%
  \global\let\svgscale\undefined%
  \makeatother%
  \begin{picture}(1,0.81449275)%
    \lineheight{1}%
    \setlength\tabcolsep{0pt}%
    \put(0,0){\includegraphics[width=\unitlength,page=1]{model_both_svg-tex.pdf}}%
    \put(0.23188406,0.54202899){\color[rgb]{0,0,0}\makebox(0,0)[t]{\lineheight{1.25}\smash{\begin{tabular}[t]{c}Linear (768)\end{tabular}}}}%
    \put(0,0){\includegraphics[width=\unitlength,page=2]{model_both_svg-tex.pdf}}%
    \put(0.76521739,0.54202899){\color[rgb]{0,0,0}\makebox(0,0)[t]{\lineheight{1.25}\smash{\begin{tabular}[t]{c}Linear (768)\end{tabular}}}}%
    \put(0,0){\includegraphics[width=\unitlength,page=3]{model_both_svg-tex.pdf}}%
    \put(0.76521739,0.71594203){\color[rgb]{0,0,0}\makebox(0,0)[t]{\lineheight{1.25}\smash{\begin{tabular}[t]{c}FlauBERT (768)\end{tabular}}}}%
    \put(0,0){\includegraphics[width=\unitlength,page=4]{model_both_svg-tex.pdf}}%
    \put(0.48985507,0.36811594){\color[rgb]{0,0,0}\makebox(0,0)[t]{\lineheight{1.25}\smash{\begin{tabular}[t]{c}Linear (520)\end{tabular}}}}%
    \put(0,0){\includegraphics[width=\unitlength,page=5]{model_both_svg-tex.pdf}}%
    \put(0.48985507,0.25217391){\color[rgb]{0,0,0}\makebox(0,0)[t]{\lineheight{1.25}\smash{\begin{tabular}[t]{c}Linear (256)\end{tabular}}}}%
    \put(0,0){\includegraphics[width=\unitlength,page=6]{model_both_svg-tex.pdf}}%
    \put(0.48985507,0.13623188){\color[rgb]{0,0,0}\makebox(0,0)[t]{\lineheight{1.25}\smash{\begin{tabular}[t]{c}Linear (2)\end{tabular}}}}%
    \put(0,0){\includegraphics[width=\unitlength,page=7]{model_both_svg-tex.pdf}}%
    \put(0.48985507,0.02028986){\color[rgb]{0,0,0}\makebox(0,0)[t]{\lineheight{1.25}\smash{\begin{tabular}[t]{c}Sigmoid\end{tabular}}}}%
    \put(0,0){\includegraphics[width=\unitlength,page=8]{model_both_svg-tex.pdf}}%
    \put(0.23188406,0.71594203){\color[rgb]{0,0,0}\makebox(0,0)[t]{\lineheight{1.25}\smash{\begin{tabular}[t]{c}Wav2Vec2 (768)\end{tabular}}}}%
  \end{picture}%
\endgroup%